# Simulation and design of shaped pulses beyond the piecewise-constant approximation

Uluk Rasulov[1], Anupama Acharya[1],
Marina Carravetta[1], Guinevere Mathies[2],
Ilya Kuprov[1,*]

[1]*School of Chemistry, University of Southampton, United Kingdom.*
[2]*Department of Chemistry, University of Konstanz, Germany.*

*i.kuprov@soton.ac.uk

## Abstract
Response functions of resonant circuits create ringing artefacts if their input changes rapidly. When physical limits of electromagnetic spectroscopies are explored, this creates two types of problems. Firstly, simulation: the system must be propagated accurately through every response transient, this may be computationally expensive. Secondly, optimal control: circuit response must be taken into account; it may be advantageous to design pulses that are resilient to such distortions. At the root of both problems is the popular piecewise-constant approximation for control sequences in the rotating frame; in magnetic resonance it has persisted since the earliest days and has become entrenched in the commercially available hardware. In this paper, we report an implementation and benchmarks of recent Lie-group methods that can efficiently simulate and optimise smooth control sequences.

# 1. Introduction

The original formulation [1] and many subsequent refinements [2-4] of the gradient ascent pulse engineering (GRAPE) method for quantum optimal control use the piecewise-constant approximation for the control Hamiltonian in the interaction representation. GRAPE also makes an unstated assumption that hardware response functions (of amplifiers, lasers, cavities, *etc.*) create negligible distortions in the control sequence [1]. When this approximation holds, for example in liquid state nuclear magnetic resonance (NMR) [5,6] and atom interferometry [7,8], optimal control theory yields impressive results. Some instrumental effects, such as radiofrequency coil wire distance modulation in magic angle spinning NMR [9,10], can be accounted for within the piecewise-constant approximation. However, there are cases where – in our hands, and in the unpublished experience of other groups – theoretically optimal piecewise-constant GRAPE control sequences inexplicably fail to generate the intended dynamics in experimental systems.

One likely reason is illustrated in Figure 1 for a typical composite radiofrequency pulse used in $^{14}$N NMR spectroscopy. The significant distortion introduced by the probe circuit [10] prevents the control sequence from taking the ensemble of $^{14}$N spins to the intended destination state. It also makes the simulation harder because the spin system sees the distorted waveform: time discretisation requirements are more stringent for the circuit output in Figure 1 compared to the input.

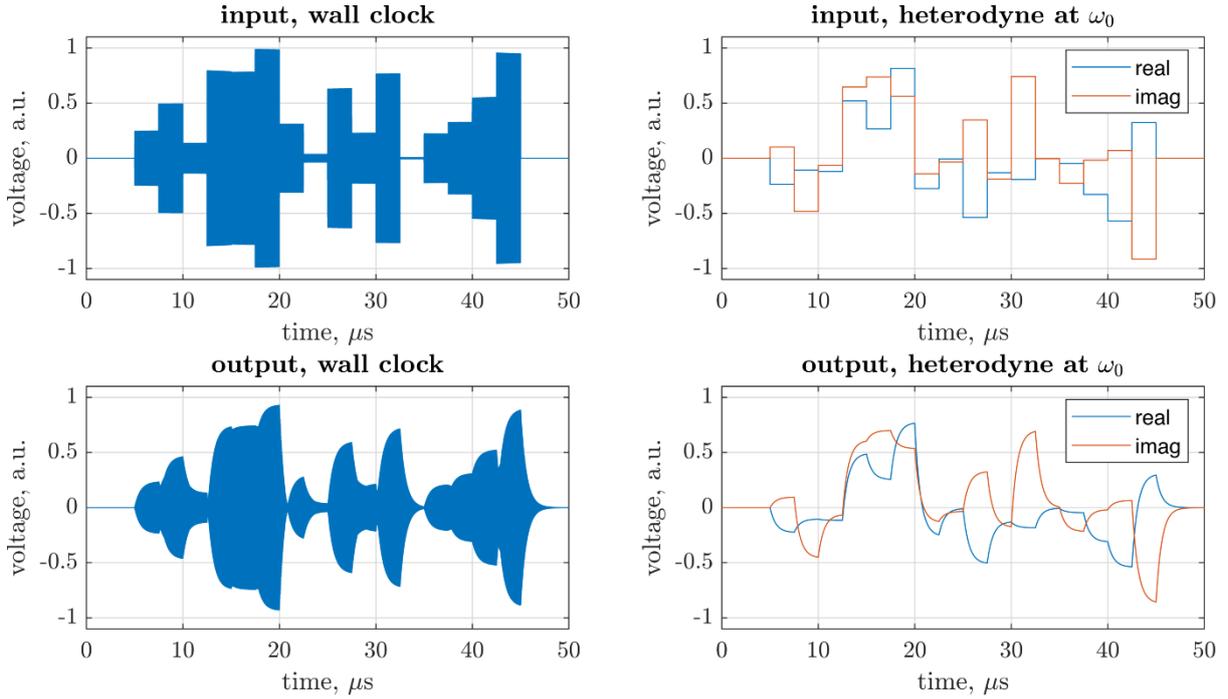

***Figure 1.*** *Distortions created in a composite pulse by a nuclear magnetic resonance probe tuned to the $^{14}$N spin precession frequency in a 14.09 Tesla magnet. The probe is modelled, using Spinach 2.8 [11], as a lumped RLC circuit with a quality factor Q = 80.* ***Top Left:*** *input voltage as a function of time; oscillations at 43.37 MHz appear as solid blocks.* ***Top Right:*** *rotating frame representation of the same pulse, indicating the in-phase (blue line) and out-of-phase (red line) components relative to the $\omega_0/2\pi$ = 43.37 MHz reference frequency.* ***Bottom Row:*** *same quantities plotted after the application of the response function of the probe RLC circuit.*

Such distortions can sometimes be ignored: they diminish for higher frequencies and smaller quality factors, they may also be tolerated because the definition of a GRAPE optimum is zero derivative (meaning first-order resilience) of the fidelity with respect to small variations in the control sequence [1]. Still, there are cases – notably, quadrupolar NMR and time-domain EPR spectroscopy [12] – where anecdotal evidence indicates that instrument response functions destroy the fidelity advantages predicted by idealised GRAPE optimisations. Figure 1 suggests that the reason is the presence of circuit response transients at the edges of the pulse sequence. An obvious solution would be to create a sequence without such edges: to move from a piecewise-constant to at least piecewise-linear sequence in the rotating frame, with appropriate upgrades to the hardware.

## 2. Two- and three-point product integrators

Upgrading the shaped pulse model from piecewise-constant to piecewise-linear in the interaction representation strikes down the first approximation made in foundational texts on numerical magnetic resonance simulation [13,14] and optimal control [1,15] – time slicing followed by the assumption that the Hamiltonian does not change within a sufficiently thin slice.

Mathematically speaking, we would be moving away from the picture where the time-ordered exponential [16] expression for the adjoint representation (*aka* "Liouville space") propagator $\mathcal{P}(t)$:

$$\frac{d}{dt}\boldsymbol{\rho}(t) = -i\mathcal{L}(t)\boldsymbol{\rho}(t), \qquad \mathcal{L}(t) = \mathcal{H}(t) + i\mathcal{R} + \ldots$$

$$\boldsymbol{\rho}(t) = \mathcal{P}(t)\boldsymbol{\rho}(0), \quad \mathcal{P}(t) = \overleftarrow{\exp}\left(-i\int_0^t \mathcal{L}(t)dt\right) \tag{1}$$

is approximated by a product integral – as the zero slice width limit of a time-ordered product:

$$\overleftarrow{\exp}\left(-i\int_0^t \mathcal{L}(t)dt\right) = \lim_{\Delta t_k \to 0} \overleftarrow{\prod_k} e^{-i\mathcal{L}(t_k)\Delta t_k} \tag{2}$$

where $\boldsymbol{\rho}(t)$ is the density matrix, $\mathcal{L}(t)$ is the evolution generator (commonly called the Liouvillian), $\mathcal{H}(t)$ is the Hamiltonian commutation superoperator, $\mathcal{R}$ is the relaxation superoperator (not usually time-dependent), arrow over the exponential indicates Dyson time order [16], slices of duration $\Delta t_k$ are centred around $t_k$, and the product is ordered from right to left in time.

### 2.1 Lie group methods

Computationally efficient extensions of this scheme to higher accuracy orders with respect to $\Delta t_k$ are recent [17-19]. They go under the general name of Lie group methods [18]; their key feature is that the approximation is made at the level of the generator of the exponential map. Consider the general algebraic form of the Lie equation in the magnetic resonance notation:

$$\frac{d}{dt}\boldsymbol{\rho}(t) = -i\mathcal{L}(t,\boldsymbol{\rho})\boldsymbol{\rho}(t) \tag{3}$$

where $\boldsymbol{\rho}$ is the state vector evolving under the (possibly dissipative, as well as time- and state-dependent) generator $\mathcal{L}$. Popular numerical methods for solving this equation (for example, Runge-Kutta [20,21]) do not observe essential conservation laws; they also fail on critical accuracy requirements (for example, trajectory endpoint phase) of quantum dynamics simulations.

It has been recognised for some time [17-19] that one class of numerical methods that does deliver on those requirements is based on the following reformulation of the time evolution problem:

$$\boldsymbol{\rho}(t) = \exp\{\boldsymbol{\Omega}(t)\}\boldsymbol{\rho}(0)$$

$$\frac{d\boldsymbol{\Omega}}{dt} = -i\sum_{m=0}^{\infty}\frac{B_m}{m!}\underbrace{[\boldsymbol{\Omega},[\boldsymbol{\Omega},...[\boldsymbol{\Omega},\mathcal{L}]]]}_{m}, \qquad \boldsymbol{\Omega}(0) = \boldsymbol{0} \quad (4)$$

where $B_k$ are Bernoulli numbers [22]. The series in the differential equation for $\boldsymbol{\Omega}(t)$ may be truncated and the equation then solved using standard numerical ODE methods [17-19]. The exponential action ensures that group-theoretical invariants (and therefore conservation laws) are observed.

Using Magnus expansions [19,23] yields the following one-point, two-point, and three-point propagation rules when the evolution generator $\mathcal{L}$ is not state-dependent:

$$\begin{aligned}
&\boldsymbol{\rho} \leftarrow \exp\{-i\mathcal{L}_L \Delta t\}\boldsymbol{\rho}, \qquad \boldsymbol{\rho} \leftarrow \exp\{-i\mathcal{L}_M \Delta t\}\boldsymbol{\rho} \\
&\boldsymbol{\rho} \leftarrow \exp\left\{-i\left(\frac{\mathcal{L}_L + \mathcal{L}_R}{2} + \frac{i}{6}[\mathcal{L}_L, \mathcal{L}_R]\Delta t\right)\Delta t\right\}\boldsymbol{\rho} \\
&\boldsymbol{\rho} \leftarrow \exp\left\{-i\left(\frac{\mathcal{L}_L + 4\mathcal{L}_M + \mathcal{L}_R}{6} + \frac{i[\mathcal{L}_L, \mathcal{L}_R]\Delta t}{12}\right)\Delta t\right\}\boldsymbol{\rho}
\end{aligned} \quad (5)$$

where L, R, and M subscripts indicate the left edge, the right edge, and the midpoint of the $\Delta t$ interval; the one-point rule is identical to the established practice in Eqs (1) and (2). These rules are different from those recently explored in the optimal control context by Dalgaard and Motzoi [24]: discretisation point locations (interval edges and midpoint instead of Gauss-Legendre quadrature points) are here dictated by the logistics of magnetic resonance hardware. Logistically, the left point of each interval is the right point of the preceding one: generators may be re-used.

The same propagation rules are also applicable to isospectral flow problems (colloquially called "Hilbert space evolution" in magnetic resonance [25]) where the propagation step involves two-sided multiplication of the density matrix [19] by the propagator and its Hermitian conjugate.

When the evolution generator does depend on the state (radiation damping [26], relaxation theories at low temperature [27], non-linear chemical kinetics [28], *etc.*), the simplest (of many possibilities [17-19]) second order Lie group method estimates the generator at the midpoint, and uses the estimate to propagate:

$$\begin{aligned}
\mathcal{L}_L &\leftarrow \mathcal{L}(t_L, \boldsymbol{\rho}_L) \\
\boldsymbol{\rho}_M &\leftarrow \exp\left(-\tfrac{i}{2}(t_R - t_L)\mathcal{L}_L\right)\boldsymbol{\rho}_L \\
\mathcal{L}_M &\leftarrow \mathcal{L}(t_M, \boldsymbol{\rho}_M) \\
\boldsymbol{\rho}_R &\leftarrow \exp\left(-i(t_R - t_L)\mathcal{L}_M\right)\boldsymbol{\rho}_L
\end{aligned} \quad (6)$$

The simplest fourth order method proceeds to estimate the generator at the right edge and adds the following stages to Eq (6):

$$\mathscr{L}_R \leftarrow \mathscr{L}(t_R, \boldsymbol{\rho}_R)$$
$$\boldsymbol{\rho}_R \leftarrow \exp\left\{-i\left(\frac{\mathscr{L}_L + 4\mathscr{L}_M + \mathscr{L}_R}{6} + \frac{i[\mathscr{L}_L, \mathscr{L}_R]\Delta t}{12}\right)\Delta t\right\}\boldsymbol{\rho}_L \quad (7)$$

Other methods in this class differ in the details of the approximations used to solve Eq (4) for $\boldsymbol{\Omega}(t)$; a comprehensive review was published by Iserles, Munthe-Kaas, Nørsett, and Zanna [18].

2.2 Numerically efficient implementations

When the propagator is computed explicitly, the additional cost of using two-and three-point rules in Eq (5) – two sparse matrix-matrix multiplications – is negligible relative to the cost of the subsequent calculation of the matrix exponential. However, computing the exponential explicitly (cubic complexity with matrix dimension and much increased memory utilisation relative to storing only the generator) is rarely efficient, particularly in situations when the generator is defined implicitly, for example as a polyadic object with un-opened Kronecker products [29] or a DMRG-type tensor structure [30]. In those cases, only the product of $\mathscr{L}$ with a user-specified vector is available, and Krylov type propagation algorithms [31] must be used – those are free of matrix-matrix multiplications. A simple illustration is the Taylor expansion of the action by the exponential of a matrix $\mathbf{A}$ on a vector $\mathbf{v}$:

$$\exp(\mathbf{A})\mathbf{v} = \left[\sum_{k=0}^{\infty} \mathbf{A}^k/k!\right]\mathbf{v} = \sum_{k=0}^{\infty} \frac{1}{k!}\mathbf{A}(\mathbf{A}\cdots(\mathbf{A}\mathbf{v})) \quad (8)$$

where the right hand side is reordered to have only matrix-vector products. Those are cheaper (quadratic complexity with dimension) than matrix-matrix products involved in the Taylor series for the exponential of $\mathbf{A}$. This simple trick is in practice equivalent to Krylov methods, where the problem is projected into the Krylov subspace of $\mathbf{A}$ and $\mathbf{v}$ spanned by the set of products $\{\mathbf{v}, \mathbf{A}\mathbf{v}, \mathbf{A}^2\mathbf{v}, ...\}$, the product $\exp(\mathbf{A})\mathbf{v}$ is computed inside the Krylov subspace, and then projected back [31]. Eq (8) is a shortcut because the expansion coefficients of $\exp(\mathbf{A})\mathbf{v}$ in $\{\mathbf{v}, \mathbf{A}\mathbf{v}, \mathbf{A}^2\mathbf{v}, ...\}$ are already known.

Technical details are given in Chapter 4 of IK's book on the subject [32]. The simplicity of adapting this process for the two- and three-point rules in Eq (5) is illustrated by *Matlab* code of the inner loop in the summation of Eq (8). The midpoint code, where `k` is the summation variable, `t/nsteps` is the subdivided time step, `L` is the Liouvillian, and `next_term` refers to the terms of the summation, is:

```
% Centre point propagator
next_term=-(1i/k)*(t/nsteps)*(L*next_term);
```

The two-point propagator adaptation of this code pre-computes $\mathscr{L}_L\boldsymbol{\rho}$ and $\mathscr{L}_R\boldsymbol{\rho}$, and re-uses them within the commutator; the overall complexity ends up being four matrix-vector multiplications per term of the Taylor series for the middle row of Eq (5):

```
% Re-usable intermediates
rho_a=L{1}*next_term; rho_b=L{2}*next_term;

% Left edge + right edge two-point propagator
next_term=-(1i/2)*(1/k)*(t/nsteps)*(rho_a+rho_b)+...
          (1/6)*(1/k)*(t^2/nsteps)*(L{1}*rho_b-L{2}*rho_a);
```

The three-point propagator in the bottom row of Eq (5), when similarly implemented, requires five matrix-vector multiplications per summation term. When the time step is appropriately scaled [32], the series converges quickly and monotonically in the norm. The implementation for the state-dependent generator case in Eqs (6) and (7) is similar, open-source code is available in *Spinach* [11].

### 2.3 Practical accuracy benchmarks

Even before their benefits in the context of optimal control theory are evaluated, the improvement in the accuracy (relative to the currently dominant piecewise-constant implementations) is so significant (Figure 2) that we recommend adopting these methods in all magnetic resonance simulation packages. Dalgaard and Motzoi were pessimistic in their conclusions section [24] about the logistical overhead of high-rank Lie group integrators, but a simple three-point integrator does actually exist (last row of Eq (5) and yellow curve in Figure 2); further hope is offered by commutator-free methods that are currently being explored [33].

For a popular Veshtort-Griffin band-selective shaped pulse [34] in NMR spectroscopy, the accuracy of the three-point integrator at 50 time slices exceeds the piecewise-constant approximation at 1000 slices. This 20-fold saving factor pertains directly to Hilbert space evolution; for Liouville space Krylov propagation, it is reduced to 4-fold on the wall clock when we take into account the greater numerical cost of the three-point propagation step, but that is still a major efficiency improvement.

Similar accuracy scaling was obtained for propagation through a rotor period of a magic angle spinning NMR simulation (Figure S1 in the Supplementary Information) and for the calculation of the effective Hamiltonian over the period of the radiofrequency rotating frame in a quadrupolar overtone NMR simulation (Figure S2 in the SI). In general, this scaling behaviour is to be expected whenever the evolution generator is time-dependent.

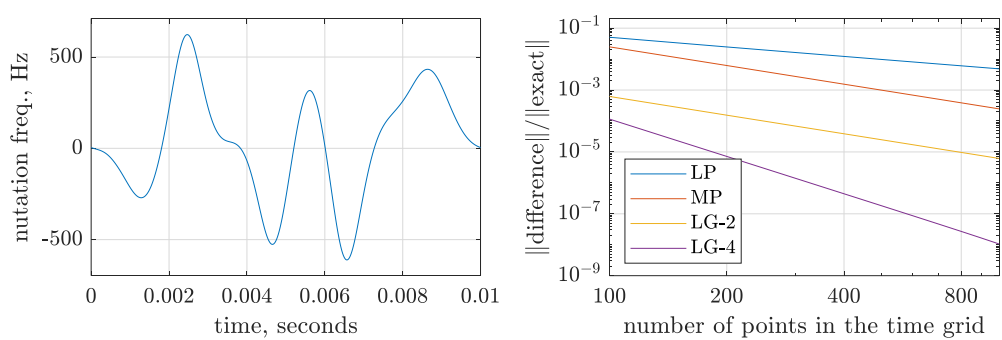

*Figure 2. (Left)* Veshtort-Griffin E1000B band-selective radiofrequency pulse used in NMR spectroscopy to achieve selective magnetisation excitation [34], here in the ±500 Hz interval for a chain of 31 J-coupled $^1$H nuclei with Larmor frequencies uniformly spaced in the ±2500 Hz interval. *(Right)* Final state accuracy as a function of discretisation point count using: (blue line, marked LP) left point piecewise-constant approximation; (red line, marked MP) midpoint piecewise-constant approximation; (yellow line, marked LG-2) second-order Lie integrator from Eq (5); (violet line, marked LG-4) fourth-order Lie integrator from Eq (5). Reproduced from the example set of Spinach library [11].

The same improvement is seen in simulations involving state-dependent generators; a common example in magnetic resonance is radiation damping, for which the modified Bloch equations are [26]:

$$\frac{\partial}{\partial t}\begin{pmatrix}\mu_X\\ \mu_Y\\ \mu_Z\end{pmatrix}=\begin{pmatrix}-r_2 & -\omega & 0\\ \omega & -r_2 & 0\\ 0 & 0 & -r_1\end{pmatrix}\begin{pmatrix}\mu_X\\ \mu_Y\\ \mu_Z-\mu_{eq}\end{pmatrix}-k_{rd}\begin{pmatrix}\mu_X\mu_Z\\ \mu_Y\mu_Z\\ \mu_X^2+\mu_Y^2\end{pmatrix} \quad (9)$$

where $\mu_{XYZ}$ are cartesian components of the magnetisation vector, $\omega$ is the Larmor frequency, $r_{1,2}$ are longitudinal and transverse relaxation rates, and $k_{rd}$ is the radiation damping rate constant. Rewriting this equation in pseudolinear form exposes the state-dependent generator:

$$\frac{\partial}{\partial t}\begin{pmatrix}\mu_X\\ \mu_Y\\ \mu_Z\end{pmatrix}=-\begin{pmatrix}r_2+k_{rd}\mu_Z & \omega & 0\\ -\omega & r_2+k_{rd}\mu_Z & 0\\ k_{rd}\mu_X & k_{rd}\mu_Y & r_1\end{pmatrix}\begin{pmatrix}\mu_X\\ \mu_Y\\ \mu_Z\end{pmatrix} \quad (10)$$

which goes into Eqs (6) and (7). The accuracy profile is shown in Figure 3 – the scaling of the residual error is the same as it was for the state-independent generator in the right panel of Figure 2.

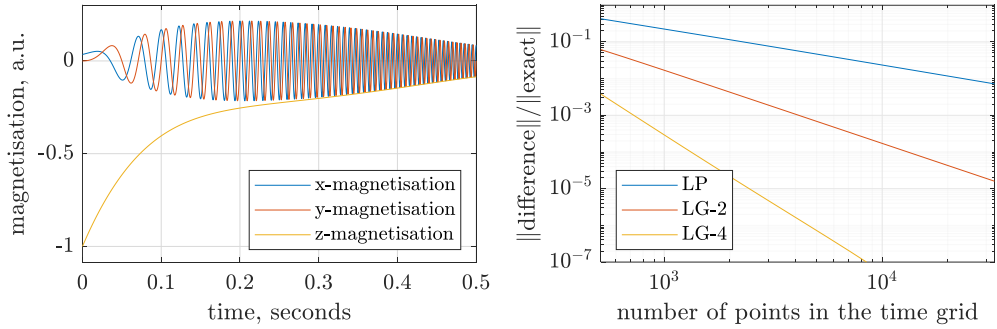

*Figure 3. (Left)* Rotating frame picture of the radiation damping during precession of an ensemble of magnetic dipoles in the presence of a linearly swept magnetic field (Zeeman frequency ramp from 0 to 200 Hz in 0.5 seconds), decoherence ($T_1 = T_2 = 0.1$ seconds), and radiation damping coefficient $k_{rd} = 40$ Hz in Eq (9). The initial condition is off by 2 degrees from the negative direction of the Z axis. *(Right)* Relative error in the final state after propagation to $t = 0.5$ s: (blue line, marked LP) left point evolution generator followed by the piecewise-constant approximation; (red line, marked LG-2) second-order Lie-group integrator from Eq (6); (yellow line, marked LG-4) fourth-order Lie-group integrator from Eq (7). Reproduced from the example set of Spinach library [11].

## 3. Lie group extension of the GRAPE algorithm

Gradient ascent pulse engineering [1-3] is an open-loop quantum control framework that seeks to maximise measures of experiment fidelity with respect to instrumentally constrained control parameters. For common dissipative ensemble control problems, the equation of motion is:

$$\frac{d}{dt}\boldsymbol{\rho}(t)=-i\mathcal{L}(t)\boldsymbol{\rho}(t),\qquad \mathcal{L}(t)=\mathcal{H}(t)+i\mathcal{R} \quad (11)$$

where $\boldsymbol{\rho}$ is the ensemble state vector (typically a vectorised density matrix [25]), $\mathcal{H}(t)$ is the unitary evolution generator (typically the Hamiltonian commutation superoperator), and $\mathcal{R}$ is the dissipative dynamics generator. The Liouvillian $\mathcal{L}(t)$ may be split into the uncontrollable "drift" part $\mathcal{D}(t)$ and a linear combination of the operators $\mathcal{C}_k$ whose coefficients $c^{(k)}(t)$ the instrument can vary:

$$\mathcal{L}(t)=\mathcal{D}(t)+\sum_k c^{(k)}(t)\mathcal{C}_k \quad (12)$$

For an experiment of duration $T$ with initial condition $\boldsymbol{\rho}_0$ and the desired destination state $\boldsymbol{\delta}$, popular measures of fidelity are functions of the overlap between $\boldsymbol{\delta}$ and $\boldsymbol{\rho}(T)$:

$$f = \langle \boldsymbol{\delta} | \boldsymbol{\rho}(T) \rangle = \langle \boldsymbol{\delta} | \overleftarrow{\exp} \left[ -i \int_0^T \mathcal{L}(t) dt \right] | \boldsymbol{\rho}_0 \rangle \tag{13}$$

where the arrow indicates Dyson's time-ordered exponential [16]. When this quantity and its variational derivatives with respect to the control sequences $c^{(k)}(t)$ are available from a simulation, the fidelity may be optimised. GRAPE does this by discretising time and assuming both the drift and the control sequences to be piecewise-constant (Figure 4, left panel):

$$c^{(k)}(t) = c_n^{(k)}, \quad \mathcal{D}(t) = \mathcal{D}_n \\ t_{n-1} < t < t_n \tag{14}$$

This yields a remarkably efficient gradient evaluation algorithm reminiscent of backpropagation [35] in machine learning, wherein the gradient of the fidelity

$$f = \langle \boldsymbol{\delta} | \mathcal{P}_N \cdots \mathcal{P}_{n+1} \mathcal{P}_n \mathcal{P}_{n-1} \cdots \mathcal{P}_1 | \boldsymbol{\rho}_0 \rangle \\ \frac{\partial f}{\partial c_n^{(k)}} = \langle \boldsymbol{\delta} | \mathcal{P}_N \cdots \mathcal{P}_{n+1} \frac{\partial \mathcal{P}_n}{\partial c_n^{(k)}} \mathcal{P}_{n-1} \cdots \mathcal{P}_1 | \boldsymbol{\rho}_0 \rangle \\ \mathcal{P}_n = \exp\left[ -i \left( \mathcal{D}_n + \sum_k c_n^{(k)} \mathcal{C}_k \right) \Delta t_n \right] \tag{15}$$

is calculated by one forward trajectory calculation from $\boldsymbol{\rho}_0$, one backward trajectory calculation with Hermitian conjugate generators from $\boldsymbol{\delta}$, and a number of inner products with propagator derivatives in the middle [5]. The efficiency extends to arbitrary waveform basis sets; this is useful when instrumental constraints exist on which control functions can be generated. Consider a real orthonormal basis set of discretised waveforms $\{\mathbf{w}_1, \ldots, \mathbf{w}_M\}$:

$$\mathbf{W} = \begin{bmatrix} | & \cdots & | \\ \mathbf{w}_1 & \cdots & \mathbf{w}_M \\ | & \cdots & | \end{bmatrix}, \quad \mathbf{W}^\mathrm{T} \mathbf{W} = \mathbf{1} \tag{16}$$

such that the control sequence $\mathbf{c}^{(k)}$ at $k$-th channel has an expansion

$$\mathbf{c}^{(k)} = \sum_m \alpha_m^{(k)} \mathbf{w}_m = \mathbf{W} \boldsymbol{\alpha}^{(k)} \quad \Leftrightarrow \quad c_n^{(k)} = \sum_m w_{nm} \alpha_m^{(k)} \tag{17}$$

Then derivatives of any function $f$ of $\mathbf{c}^{(k)}$ vector are translated as follows into the corresponding derivatives with respect to the waveform basis expansion coefficient vector $\boldsymbol{\alpha}^{(k)}$:

$$\frac{\partial f}{\partial \alpha_m^{(k)}} = \sum_n \frac{\partial f}{\partial c_n^{(k)}} \frac{\partial c_n^{(k)}}{\partial \alpha_m^{(k)}} = \sum_n \frac{\partial f}{\partial c_n^{(k)}} w_{nm} \quad \Leftrightarrow \quad \nabla_{\boldsymbol{\alpha}} f = \mathbf{W}^\mathrm{T} [\nabla_{\mathbf{c}} f] \\ \frac{\partial^2 f}{\partial \alpha_m^{(k)} \partial \alpha_{m'}^{(k')}} = \sum_{n,n'} \frac{\partial f}{\partial c_n^{(k)} \partial c_{n'}^{(k')}} w_{nm} w_{n'm'} \tag{18}$$

where $m$ enumerates basis waveforms, $n$ enumerates time points and $k$ enumerates control channels. These are special cases of matrix calculus chain rules; these relations connect the methods that use waveform basis sets to GRAPE algorithms [1,2].

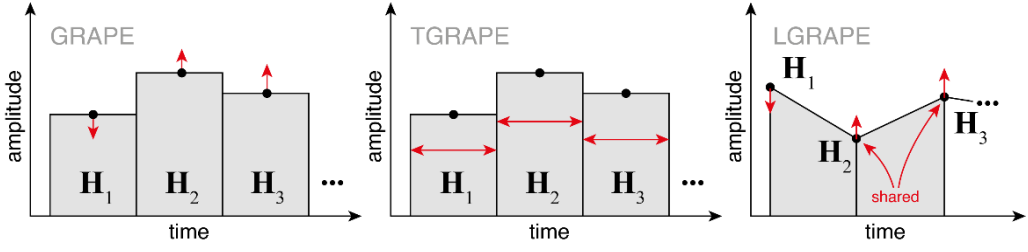

*Figure 4. Schematic illustrations of the control sequence parameter update stage in GRAPE [1] (variation of piece-wise-constant control coefficients, **left**), TGRAPE (Section 3.2, variation of slice durations with piecewise-constant control coefficients, **middle**), and of the extension of the same principle to continuous control sequences (Sections 2.1 and 3.1, piecewise-linear version shown, abbreviated LGRAPE, **right**).*

Eq (15) essentially relies on the control sequence being piecewise-constant; the shortcomings of this approach are described in the Introduction. Lifting this constraint could alleviate the hardware response function problem – the control sequence would no longer be piecewise-constant – but it would also require the corresponding update to the mathematics and the logistics of Eq (15) because adjacent Hamiltonians would influence the evolution in more than one slice (Figure 4, right panel).

### 3.1 Control sequence derivatives

Let us now apply the explicit assumption that the control sequence generated by the instrument is not piecewise-constant, but piecewise-linear. The most accurate three-point propagation rule in Eq (5) is simplified because now $\mathscr{L}_M = (\mathscr{L}_L + \mathscr{L}_R)/2$, and therefore:

$$\boldsymbol{\rho} \leftarrow \exp\left\{-i\left(\frac{\mathscr{L}_L + \mathscr{L}_R}{2} + \frac{i[\mathscr{L}_L, \mathscr{L}_R]\Delta t}{12}\right)\Delta t\right\}\boldsymbol{\rho} \tag{19}$$

Note the different coefficient in front of the commutator compared to the two-point rule in Eq (5): this is because we constrain the interval midpoint generator $\mathscr{L}_M$ to a specific value here. The expression for the fidelity is modified because each propagator now depends on two adjacent points of the discretised control sequence, and adjacent propagators share a point:

$$f = \langle\boldsymbol{\delta}|\mathscr{P}_{N,N-1}\mathscr{P}_{N-1,N-2}\cdots\mathscr{P}_{2,1}\mathscr{P}_{1,0}|\boldsymbol{\rho}_0\rangle, \qquad \mathscr{L}_n = \mathscr{D}_n + \sum_k c_n^{(k)}\mathscr{C}_k$$

$$\mathscr{P}_{n+1,n} = \exp\left\{-i\left(\frac{\mathscr{L}_n + \mathscr{L}_{n+1}}{2} + \frac{i}{12}[\mathscr{L}_n, \mathscr{L}_{n+1}]\Delta t\right)\Delta t\right\} \tag{20}$$

There are three types of derivatives in this setting – the first-point (one propagator), the last point (one propagator), and midpoint (two adjacent propagators):

$$\frac{\partial f}{\partial c_0^{(k)}} = \langle\boldsymbol{\delta}|\mathscr{P}_{N,N-1}\mathscr{P}_{N-1,N-2}\cdots\mathscr{P}_{2,1}\left(\frac{\partial}{\partial c_0^{(k)}}\mathscr{P}_{1,0}\right)|\boldsymbol{\rho}_0\rangle$$

$$\frac{\partial f}{\partial c_N^{(k)}} = \langle\boldsymbol{\delta}|\left(\frac{\partial}{\partial c_N^{(k)}}\mathscr{P}_{N,N-1}\right)\mathscr{P}_{N-1,N-2}\cdots\mathscr{P}_{3,2}\mathscr{P}_{1,0}|\boldsymbol{\rho}_0\rangle \tag{21}$$

$$\frac{\partial f}{\partial c_n^{(k)}} = \langle\boldsymbol{\delta}|\mathscr{P}_{N,N-1}\mathscr{P}_{N-1,N-2}\cdots\left(\frac{\partial}{\partial c_n^{(k)}}(\mathscr{P}_{n+1,n}\mathscr{P}_{n,n-1})\right)\cdots\mathscr{P}_{3,2}\mathscr{P}_{1,0}|\boldsymbol{\rho}_0\rangle$$

For the first-point and the last-point propagators, the directional derivative calculation problem is already solved [2,36]; the most elegant method uses auxiliary matrices:

$$\exp\left[\begin{pmatrix} \mathbf{A} & \partial\mathbf{A}/\partial\alpha \\ \mathbf{0} & \mathbf{A} \end{pmatrix}\right] = \begin{pmatrix} e^{\mathbf{A}} & \partial e^{\mathbf{A}}/\partial\alpha \\ \mathbf{0} & e^{\mathbf{A}} \end{pmatrix} \quad (22)$$

Numerical implementations are available in *Spinach* [11], including the cases where only the action by an exponential derivative on a vector is needed [3,36]. It bears notice (proof by induction for matrix powers, followed by Taylor series) that also:

$$f\left[\begin{pmatrix} \mathbf{A} & \partial\mathbf{A}/\partial\alpha \\ \mathbf{0} & \mathbf{A} \end{pmatrix}\right] = \begin{pmatrix} f(\mathbf{A}) & \partial f(\mathbf{A})/\partial\alpha \\ \mathbf{0} & f(\mathbf{A}) \end{pmatrix} \quad (23)$$

for any scalar function $f(x)$ that can be extended to a matrix function using Taylor series. Propagator product derivatives are reduced by the application of matrix product rule:

$$\frac{\partial}{\partial c_n^{(k)}}\left(\mathcal{P}_{n+1,n}\mathcal{P}_{n,n-1}\right) = \left(\frac{\partial}{\partial c_n^{(k)}}\mathcal{P}_{n+1,n}\right)\mathcal{P}_{n,n-1} + \mathcal{P}_{n+1,n}\left(\frac{\partial}{\partial c_n^{(k)}}\mathcal{P}_{n,n-1}\right) \quad (24)$$

The problem is therefore reduced to calculating the derivatives of the generator in the curly brackets of Eq (20) with respect to the control coefficients. Consider a particular time interval with a left (L) and a right (R) time point. The interval evolution generator is:

$$\mathfrak{L} = \frac{\mathfrak{L}_L + \mathfrak{L}_R}{2} + \frac{i}{12}[\mathfrak{L}_L, \mathfrak{L}_R]\Delta t$$
$$\mathfrak{L}_L = \mathfrak{D}_L + \sum_k c_L^{(k)}\mathcal{C}_k, \quad \mathfrak{L}_R = \mathfrak{D}_R + \sum_k c_R^{(k)}\mathcal{C}_k \quad (25)$$

where $\mathfrak{D}_{L,R}$ are drift generators, $\mathcal{C}_k$ are control operators, $c_{L,R}^{(k)}$ are control coefficients, and the sum runs over control channels. After straightforward rearrangements and differentiation, we arrive at the following expressions for the derivatives of the Liouvillian with respect to the control coefficients:

$$\frac{\partial \mathfrak{L}}{\partial c_R^{(k)}} = \left(\frac{\mathcal{C}_k}{2} + \frac{i\Delta t}{12}[\mathfrak{D}_L, \mathcal{C}_k]\right) + \frac{i\Delta t}{12}\sum_n c_L^{(n)}[\mathcal{C}_n, \mathcal{C}_k]$$
$$\frac{\partial \mathfrak{L}}{\partial c_L^{(k)}} = \left(\frac{\mathcal{C}_k}{2} + \frac{i\Delta t}{12}[\mathcal{C}_k, \mathfrak{D}_R]\right) + \frac{i\Delta t}{12}\sum_n c_R^{(n)}[\mathcal{C}_k, \mathcal{C}_n] \quad (26)$$

under the specific assumption that the control sequence is piecewise-linear at the hardware level.

### 3.2 Time slice duration derivatives

An under-appreciated optimisation strategy within the GRAPE framework is variation of time slice durations (Figure 4, middle panel). Its practical value comes from dismal instrumental realities: once a particular experiment is running, adjusting time slice durations may be the easiest thing to do.

Within the piecewise-constant original formulation of GRAPE [1], obtaining the gradient of the fidelity with respect to the slice duration vector $\boldsymbol{\tau}$ is straightforward:

$$\frac{\partial f}{\partial \tau_n} = \langle \delta | \mathcal{P}_N \cdots \mathcal{P}_{n+1} \frac{\partial \mathcal{P}_n}{\partial \tau_k} \mathcal{P}_{n-1} \cdots \mathcal{P}_1 | \rho_0 \rangle$$
$$\frac{\partial \mathcal{P}_n}{\partial \tau_k} = \frac{\partial}{\partial \tau_k} \exp[-i\mathcal{L}_n \tau_n] = -i\mathcal{L}_n \exp[-i\mathcal{L}_n \tau_n] \quad (27)$$

which is compatible with the efficient implementation logistics described above – the system is propagated forward from the initial condition, backward from the destination state, and the elements of the fidelity gradient are calculated in a parallel loop:

$$\frac{\partial f}{\partial \tau_n} = -i \langle \delta_n | \mathcal{L}_n | \rho_n \rangle, \qquad \mathcal{P}_n = \exp[-i\mathcal{L}_n \tau_n]$$
$$\langle \delta_n | = \langle \delta | \mathcal{P}_N \cdots \mathcal{P}_{n-1}, \qquad |\rho_n\rangle = \mathcal{P}_n \cdots \mathcal{P}_1 | \rho_0 \rangle \quad (28)$$

The extension to the situation when the control sequence generated by the experimental hardware is not piecewise-constant, but piecewise-linear, follows from the results of the preceding section:

$$\frac{\partial f}{\partial \tau_n} = \langle \delta | \mathcal{P}_{N,N-1} \mathcal{P}_{N-1,N-2} \cdots \frac{\partial \mathcal{P}_{n,n-1}}{\partial \tau_n} \cdots \mathcal{P}_{2,1} \mathcal{P}_{1,0} | \rho_0 \rangle$$
$$\mathcal{P}_{n,n-1} = \exp\left\{ -i \left( \frac{\mathcal{L}_{n-1} + \mathcal{L}_n}{2} + \frac{i}{12}[\mathcal{L}_{n-1}, \mathcal{L}_n] \tau_n \right) \tau_n \right\} \quad (29)$$

The derivative of this propagator with respect to slice duration is straightforward. Consider an interval of duration $\tau$ with a left edge Liouvillian $\mathcal{L}_\mathrm{L}$ and a right edge Liouvillian $\mathcal{L}_\mathrm{R}$. Eq (22) and its numerically efficient refinements [3,36] are directly applicable with:

$$\mathbf{A} = -i \left( \frac{\mathcal{L}_\mathrm{L} + \mathcal{L}_\mathrm{R}}{2} + \frac{i}{12}[\mathcal{L}_\mathrm{L}, \mathcal{L}_\mathrm{R}] \tau \right) \tau$$
$$\frac{\partial \mathbf{A}}{\partial \tau} = -i \frac{\mathcal{L}_\mathrm{L} + \mathcal{L}_\mathrm{R}}{2} + \frac{1}{6}[\mathcal{L}_\mathrm{L}, \mathcal{L}_\mathrm{R}] \tau \quad (30)$$

and likewise for any other generator in the Lie-group methods family. An open-source numerical implementation of slice-duration GRAPE is available in *Spinach* [11].

### 3.3 Practical benchmarks: broadband pulse

Consider a 90-degree universal rotation $^{13}$C pulse in a modern 28.2 Tesla (1.2 GHz proton frequency) NMR magnet. The pulse must accomplish the following transformation of the basis operators:

$$\mathbf{S}_Z \to \mathbf{S}_X, \quad \mathbf{S}_Y \to \mathbf{S}_Y, \quad \mathbf{S}_X \to -\mathbf{S}_Z \quad (31)$$

uniformly within a bandwidth of around 200 ppm (≈ 60 kHz) and must be short enough for the worst-case $^{13}$C-$^{1}$H *J*-coupling (around 200 Hz) to have a negligible effect. The latter requirement caps the pulse duration at about 1/100*J* = 50 μs. Maximum instrumentally achievable nutation frequency varies from 50 to 70 kHz across the radiofrequency coil of the NMR probe, and it is therefore clear that a hard $^{13}$C pulse (*i.e.* the shortest pulse at the maximum available RF power, here about 4 μs) is not possible in a 1.2 GHz magnet due to significant phase errors across the spectral window (Figure 5, top panel). A key advantage of optimal control theory is the ability [5] to generate pulse waveforms that are: (a) free of such errors; (b) more resistant to resonance offset and power miscalibration than composite pulses; (c) able to accommodate secondary considerations, such as keyhole subspaces and dead times [32]. Optimal control pulses are longer than hard pulses, but they still fit comfortably into the timing window imposed by *J*-couplings; an example is shown in the bottom panel of Figure 5.

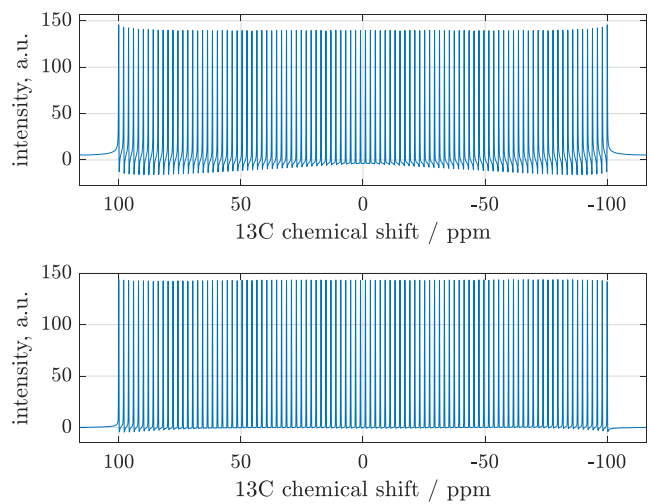

*Figure 5. The effect of universal rotation pulses intended to accomplish the state space transformation in Eq (31) for an ensemble of 100 $^{13}$C nuclei (spread uniformly over a ±100 ppm interval) in a 28.2 Tesla (1.2 GHz proton frequency) NMR magnet. **Top panel**: $^{13}$C NMR spectrum following a 4.2 µs hard pulse at 60 kHz nutation frequency, which is the limit of currently available hardware. **Bottom panel:** $^{13}$C NMR spectrum following a 40 µs piecewise-constant optimal control pulse with 400 time slices, designed to maintain the same accuracy in the nutation frequency interval between 50 and 70 kHz. See [5] for further information on such pulses; the code generating this figure is available as a part of the example set of Spinach 2.8 and later.*

In this section, we demonstrate that piecewise-linear versions of such pulses are never worse (in either convergence or performance), and in some circumstances are better than piecewise-constant versions; this is illustrated in Figure 6. The two panels on the left are "spaghetti plots", giving the infidelity as a function of LBFGS-10 iteration count for the pulse described in the caption of Figure 5: it is clear that there is no significant difference in convergence behaviour between the standard piecewise-constant LBFGS-GRAPE [2] and its piecewise-linear extension advocated in this paper; the same behaviour is observed for a few dozen other optimal control problems in the example set of *Spinach*. The panel on the right demonstrates that, for pulses with few time intervals, the piecewise-linear version of GRAPE outperforms the piecewise-constant one. For finely discretised pulses, the difference disappears because highly optimal (and parameter distribution resilient) universal rotation pulses tend to be smooth. Thus, the piecewise-linear version of GRAPE is advantageous in tightly timed experiments where the instrument only permits a small number of time slices.

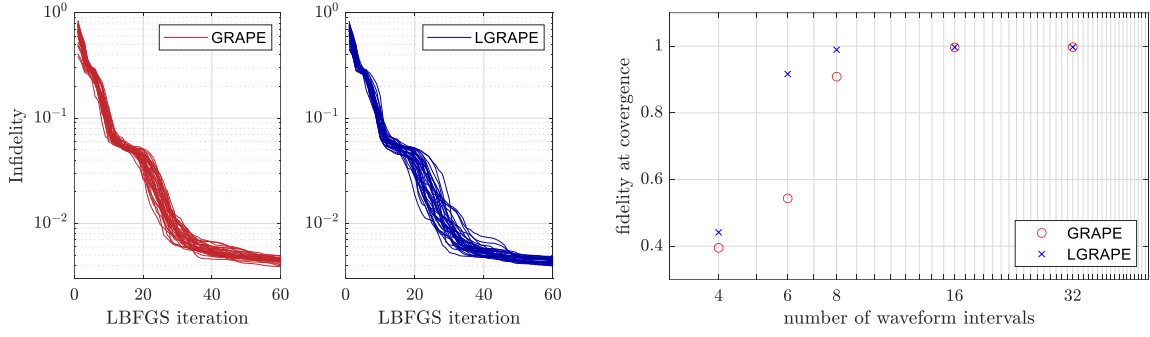

*Figure 6. Relative performance illustrations for piecewise-constant (parameterised by one point per time interval) and piecewise-linear (parameterised by two points per time interval) GRAPE algorithms for the $^{13}$C excitation pulse described in Section 3.3 of the main text. **Left and middle panel:** "spaghetti plots" illustrating LBFGS convergence behaviour of the two versions of GRAPE for a phase-modulated pulse with 60 time intervals, starting from a random initial guess. The convergence behaviour is essentially identical. **Right panel:** fidelity at convergence as a function of the number of intervals in the waveform for a phase-modulated pule with the overall duration of 51.2 µs. The more flexible piecewise-linear version of GRAPE advocated in this paper shows better performance with fewer time discretisation intervals.*

### 3.4 Practical benchmarks: prephasing pulse

An important use case for optimal control theory is dead time elimination: a control sequence may be designed to set the dynamics up for arriving at the desired destination *some time after* the controls are switched off. In magnetic resonance, this is beneficial for NMR of low-γ nuclei (because RLC circuit response effects are stronger at lower frequencies) and for quadrupolar NMR of solid powders (because of rapid ensemble dephasing by nuclear quadrupolar interaction). In both cases, the sweep width of the spectrum can be in the MHz, necessitating sub-microsecond time slices. Short time slices then render the pulses vulnerable to the distortions introduced by the hardware.

Here we model the NMR probe as a series RLC circuit; this is a rough approximation, but it is expected to capture the differences between piecewise-constant and piecewise-linear (in the rotating frame) pulse input. The transfer function is obtained from a combination of Kirchhoff's first law [37], Ohm's law [38], inductor equation, and capacitor equation [39]:

$$V_{IN}(t) = V_R(t) + V_L(t) + V_C(t)$$
$$V_R(t) = RI(t), \quad V_L(t) = LI'(t), \quad I(t) = CV'_C(t) \quad (32)$$

where primes indicate time derivatives, $V_{IN}$ is input voltage, $R$ is the resistance of the resistor, $C$ is the capacitance of the capacitor, $L$ is the inductance of the inductor, $I$ is current in the circuit, and $V_{\{R,L,C\}}$ are voltages across the resistor, inductor, and capacitor, respectively. For zero initial conditions in the Laplace domain, the corresponding equations are:

$$V_{IN}(s) = V_R(s) + V_L(s) + V_C(s)$$
$$V_R(s) = RI(s), \quad V_L(s) = sLI(s), \quad I(s) = sCV_C(s) \quad (33)$$

The quantity seen by the spin system is the magnetic field generated within the inductor, that quantity is proportional to the current. Thus, the transfer function of interest is:

$$T(s) \propto I(s)/V_{IN}(s) \propto \frac{Q^{-1}\omega_0^{-1}}{\omega_0^{-2}s^2 + Q^{-1}\omega_0^{-1}s + 1} \quad (34)$$

where $\omega_0 = 1/\sqrt{LC}$ is the resonance frequency and $Q = \sqrt{L/C}/R$ is the quality factor. In practice, each pulse was transformed from the rotating frame into the laboratory frame by mixing it with the carrier frequency $\omega_0$, then supplied to Matlab Control System Toolbox alongside the transfer function in Eq (34), and the response then heterodyned numerically back into the rotating frame. Documented source code is available (restrans.m) as a part of *Spinach* library versions 2.8 and later.

The test case was a $^2$H excitation pulse in a 14.1 T magnet, designed to set transverse deuterium magnetisation in –CD$_3$ alanine powder up for refocussing 100 μs after the end of the pulse. The quadrupole interaction tensor anisotropy (partially averaged by rapid methyl group rotation in the room temperature solid) is around 40 kHz [40]. A pulse with 24 equal time slices, overall duration of 156 μs and the maximum nutation frequency of 35 kHz per channel was optimised with the first two and the last two pulse discretisation points frozen at zero to comply with the response theory assumptions in Eqs (32)-(34). The optimisation was carried out simultaneously for 200 uniformly distributed orientations obtained from the REPULSION procedure [41] using its implementation in *Spinach* [11]. RLC distortions with $Q = 200$ and $\omega_0$ set to deuterium Larmor frequency were then applied to both waveforms.

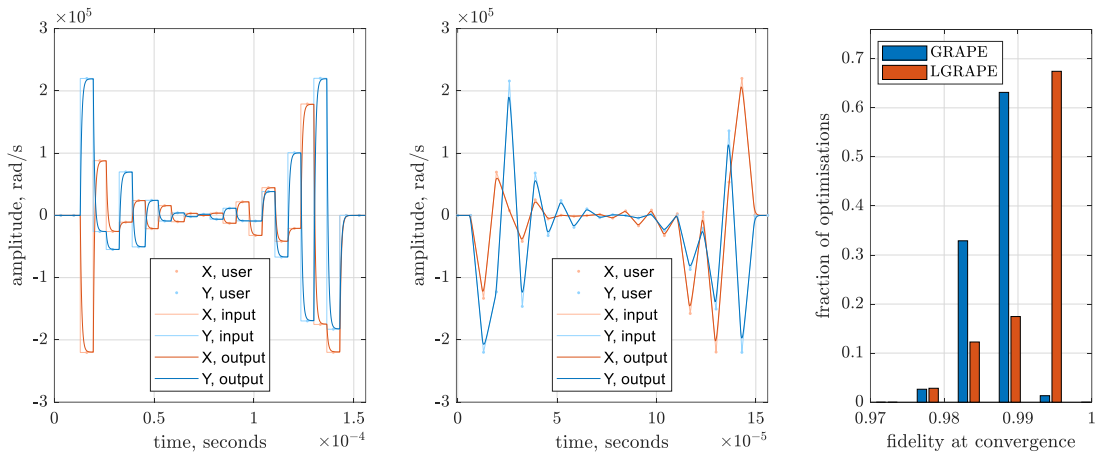

*Figure 7. An illustration of the fact that piecewise-constant composite NMR pulses suffer greater RLC circuit distortions and yield smaller fidelities at convergence than piecewise-linear pulses. Faint orange and blue lines indicate RLC circuit inputs, strong lines indicate the outputs after the RLC distortion with Q=200 is applied. **Left panel:** a conventional piecewise-constant NMR pulse, designed using GRAPE [1] to start with longitudinal magnetisation and produce perfect refocussing of transverse $^2$H magnetisation in –CD$_3$ alanine powder 100 μs after the end of the pulse. **Middle panel:** a piecewise-linear pulse, designed using the method proposed in this paper and plotted with RLC time delay compensation, accomplishing the same objective. **Right panel:** fidelity-at-convergence histograms for 200 pulses optimised from random initial guesses using 1-point and 2-point versions of the GRAPE algorithm.*

Several hundred piecewise-constant (Figure 7, left panel) and piecewise-linear (Figure 7, middle panel) pulses were obtained by starting the optimisation from different random initial guesses. It is clear from Figure 7 that piecewise-linear pulses suffer less from RLC circuit distortions and yield higher fidelities at convergence than piecewise-constant ones. In our hands (however, see also [24]), the difference in performance between the two types of pulses is not dramatic, and only manifests for tightly timed pulses accomplishing difficult objectives.

## 5. Conclusions

The historically dominant piecewise-constant approximation for shaped magnetic resonance pulses was in some ways the worst possible choice. From the mathematical point of view, it corresponds to the lowest accuracy Lie group integrator; from the engineering side, it creates the strongest RLC circuit response transients. What the sample actually sees is the distorted pulse. Experimentally, this means that each module of a magnetic resonance instrument should now ideally come with a documented response function. From the simulation point of view, it has been clear for some time [18,19,24,42] that Lie-group integrators provide dramatically greater accuracy with only a modest increase in logistical and computational costs; here we have added a number of practical efficiency improvements in the context of quantum optimal control and NMR spectroscopy, and reported an implementation in the open-source *Spinach* library [11] for spin dynamics simulations.

This implementation has allowed us to extend the GRAPE framework [1-3] for quantum optimal control to situations when the control sequences generated by experimental hardware are not piecewise-constant, but piecewise-linear in the rotating frame, and therefore more resilient to distortions introduced by the response functions of experimental hardware. This extension, including integration with LBFGS, ensemble control, and dissipative control, is available in versions 2.8 and later of *Spinach*.

## Acknowledgements

This work was supported by EPSRC (EP/W020343/1) and MathWorks, and used NVIDIA Tesla A100 GPUs through NVIDIA Academic Grants Programme. We are grateful to Sergio Blanes, Fernando Casas and Arieh Iserles for useful discussions, and to Jos Martin, Raymond Norris and Alison Eele at MathWorks for expert technical support with Parallel Matlab. The authors acknowledge the use of the IRIDIS High Performance Computing Facility, and associated support services at the University of Southampton, in the completion of this work.

SUPPLEMENTARY INFORMATION

FOR

# Simulation and design of shaped pulses beyond the piecewise-constant approximation

Uluk Rasulov[1], Anupama Acharya[1],
Marina Carravetta[1], Guinevere Mathies[2],
Ilya Kuprov[1,*]

[1]*School of Chemistry, University of Southampton, United Kingdom.*
[2]*Department of Chemistry, University of Konstanz, Germany.*

*i.kuprov@soton.ac.uk

## S1. Further benchmarks for Lie group methods

Accuracy benchmarks described in Section 2.3 of the main text were also preformed for propagation through a rotor period of a magic angle spinning (MAS) NMR experiment. The accuracy scaling (Figure S1) of the final state is very similar to the left panel in Figure 2 of the main text.

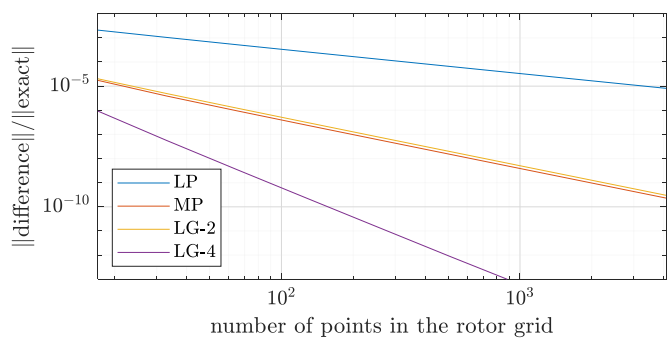

*Figure S1. Scaling of the final state error after propagation of a MAS NMR simulation through a single rotor cycle as a function of time grid point count in the rotor period, using: (blue line, marked LP) left point piecewise-constant approximation; (red line, marked MP) midpoint piecewise-constant approximation; (yellow line, marked LG-2) second-order Lie integrator from Eq (5) of the main text; (violet line, marked LG-4) fourth-order Lie integrator from Eq (5) of the main text. The system contains two protons (5.0 ppm, 2.0 ppm) in a 600 MHz NMR magnet at a distance of 3.9 Angstrom; the sample is spun at 50 kHz.*

The midpoint integrator is close in accuracy to the second order Lie integrator in Figure S1. This is likely because the time dependence of the Hamiltonian is here particularly simple – second rank spherical harmonics with constant angular velocity in the argument.

The same benchmark was also run in the context of the average Hamiltonian theory, for the error (in the matrix 2-norm) of the effective propagator over a radiofrequency period in a $^{14}$N quadrupolar overtone experiment (Figure S2). We see another instance of two methods being asymptotically close in accuracy: here second and fourth order Lie group methods show similar performance. This is likely because the time dependence in the Hamiltonian is once again simple: a cosine in tis case.

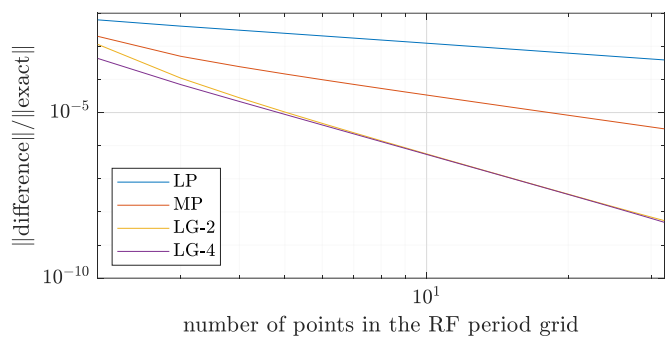

*Figure S2.* Scaling of the RF period propagator error in a $^{14}$N overtone MAS NMR simulation as a function of time grid point count in the RF period, using: (blue line, marked LP) left point piecewise-constant approximation; (red line, marked MP) mid-point piecewise-constant approximation; (yellow line, marked LG-2) second-order Lie integrator from Eq (5) of the main text; (violet line, marked LG-4) fourth-order Lie integrator from Eq (5) of the main text. The system contains a single $^{14}$N nucleus with the following quadrupolar interaction tensor parameters $C_q = 1.18 \times 10^6$ Hz, $\eta = 0.53$; the sample is spun at 20 kHz.

Both benchmarks are available in the example set of *Spinach* 2.8 and later versions [1].

## S2. References

[1] H.J. Hogben, M. Krzystyniak, G.T. Charnock, P.J. Hore, I. Kuprov, Spinach – a software library for simulation of spin dynamics in large spin systems, Journal of Magnetic Resonance, 208 (2011) 179-194.